\documentclass[a4paper,preprintnumbers,showpacs,twocolumn,superscriptaddress,nofootinbib,amsmath,amssymb]{revtex4-1}

\usepackage[dvips]{graphics}
\usepackage[hypertex]{hyperref}
\usepackage{color,hyperref}
\usepackage{times}
\usepackage{mathrsfs}
\hypersetup{colorlinks=true,linkcolor=blue,citecolor=magenta,filecolor=magenta,urlcolor=blue
}

\usepackage{enumitem}
\usepackage{graphicx,subfigure}
\usepackage{dcolumn}
\usepackage{bm}
\usepackage{color}

\def\beq{\begin{equation}}
\def\eeq{\end{equation}}
\def\bear{\begin{eqnarray}}
\def\ear{\end{eqnarray}}

\def\L{\mathscr{L}}

\begin{document}

\title{Can a light ray distinguish charge of a black hole in nonlinear electrodynamics?}

\author{Bobir Toshmatov}
\email{toshmatov@astrin.uz} \affiliation{Ulugh Beg Astronomical
Institute, Astronomy 33, Tashkent 100052, Uzbekistan}
\affiliation{Webster University in Tashkent, Alisher Navoiy 13,
Tashkent 100011, Uzbekistan} \affiliation{Tashkent Institute of
Irrigation and Agricultural Mechanization Engineers, Kori Niyoziy
39, Tashkent 100000, Uzbekistan}

\author{Bobomurat Ahmedov}
\email{ahmedov@astrin.uz} \affiliation{Ulugh Beg Astronomical
Institute, Astronomy 33, Tashkent 100052, Uzbekistan}
\affiliation{Tashkent Institute of Irrigation and Agricultural
Mechanization Engineers, Kori Niyoziy 39, Tashkent 100000,
Uzbekistan}

\author{Daniele Malafarina}
\email{daniele.malafarina@nu.edu.kz} \affiliation{Department of
Physics, Nazarbayev University, 53 Kabanbay Batyr, 010000
Nur-Sultan, Kazakhstan}

\begin{abstract}

It is a well-known fact that light rays do not follow the null
geodesics of the space-time in nonlinear electrodynamics; instead,
they follow the null geodesics of the so-called effective
space-time. Taking this into account, in this paper, we aim to
discuss the possibility of distinguishing the type of charge with
which the black hole is endowed, via the motion of light rays. The
results show that, for any black hole being a charged solution of
the field equations of general relativity coupled to the nonlinear
electrodynamics, one cannot distinguish the two types of charge
(magnetic or electric) through the motion of light rays around it.

\end{abstract}

\maketitle

\section{Introduction}\label{sec-intr}

Despite the fact that Maxwell (linear) electrodynamics and general
relativity are completely different classical theories, they both
endure the so-called singularity problem, i.e. the fact that
physically viable solutions of the theory generically exhibit
singularities. The problem is more pronounced in general
relativity, where the occurrence of curvature singularities
disrupts the causal structure of the space-time. However, ideally,
it would be preferable to develop classical theories for
gravitation and electromagnetism that do not present
singularities. It is clear that the solution of the singularity
problem requires us to extend beyond classical general relativity
and Maxwell electrodynamics, since these theories cannot avoid or
solve the problem on their own.

In electrodynamics, experiments direct us to consider the
Lagrangian density of the field to be approximately linear. In the
simplest case, it is exactly linear so that if the field equations
are solved, together with the Lorentz gauge, one will end up with
the well-known Maxwell equations~\cite{Landau-Lifshitz}. The
validity of Maxwell's equations at the classical level have been
widely demonstrated from experimental physics. However,
theoretically, if these equations are solved for point charges,
one would obtain diverging field quantities at the location of the
point charge, a fact that is already quite inexplicable, which in
turn results in infinite total energy for the electric field of a
point charge, which clearly is physically undesirable.

General relativity also allows for the existence of space-time
singularities. Of course, not all singularities appearing in
solutions of Einstein's equations are physical: For example,
coordinate singularities that are defined by the divergence of one
of the metric functions, may be merely mathematical, i.e. they can
be eliminated with an appropriate coordinate transformation.
However, curvature singularities, defined by the divergence of
curvature invariants such as the Kretschmann scalar, cannot be
eliminated by any change of coordinates and they are an intrinsic
feature of the geometry. The singularity theorems show that such
curvature singularities are an inevitable outcome of physically
viable scenarios such as, for example, the dynamical collapse that
leads to the formation of a black hole \cite{Penrose:PRL:1965},
thus making the problem of their resolution an important piece of
the hunt for a better theory of gravity and of our understanding
of extreme astrphysical phenomena \cite{Malafarina:2017}. The
existence of curvature singularities is still one of the unsolved
problems of the theory. Many approaches have been taken in the
attempt to avoid this `space-time pathology' and one of the most
promising worked-out methods is based on coupling general
relativity to nonlinear
electrodynamics~\cite{ABG:PRL:1998,ABG:PLB:2000,Bronnikov:PRL:2000,Dymnikova:CQG:2004}.
These solutions can be of electrically, magnetically or dyonically
(i.e. simultaneously electrically and magnetically) charged black
holes
\cite{Bronnikov:PRD:2001,Fan:PRD:2016,Bronnikov:PRD2017,Bronnikov:GC:2017,Toshmatov:PRD:2017,TSA:PRDnew,Kruglov:GC:2019},
like in the case of general relativity coupled to the linear
electrodynamics that results the Reissner-Nordstr\"{o}m solution.

Other new interesting phenomena appear in nonlinear
electrodynamics. One of the such phenomena is associated with the
propagation of light rays. It is a well-known fact that
electromagnetic waves propagate along null geodesics of the
space-time in vacuum and linear electrodynamics. However, this is
not the case if the electromagnetic field is self-interacting as
in the case of nonlinear electrodynamics. Then light rays do not
follow the null geodesics of a given space-time metric, instead
the paths of light can be described in terms of an effective
space-time metric which represents a modification of the original
space-time~\cite{Novello:PRD:2000,Novello:PRD:2001,ObukhovPRD:66,BretonPRD:2005,2016CQGra33h5004S,2015IJMPD2450020S,2015JCAP06:048S,2019EPJC79:988S,2019ApJ887:145S,2019ApJ874:12S,2019EPJC79:44S,2020arXiv200804118K}.
This phenomenon can also be shown from perturbations theory, as in
the high energy limit, the effective potential of the
electromagnetic perturbations of the black hole in nonlinear
electrodynamics coincides with the one governing the photon motion
in the field of a central object
\cite{SarbachPRD:67,TSSA:PRD:2018,LiEPJC:75,TSA:PRD:98b,Toshmatov:2019gxg}.

Motivated by the peculiar phenomena of nonlinear electrodynamics
discussed above, i.e. that light rays do not follow null
geodesics, in this paper we aim to determine whether it is
possible to distinguish the type of charge of the black hole from
the motion of light rays in the given geometry. To do so, we
consider a given spherically symmetric, static space-time that is
either of electrically charged or magnetically charged black hole.
By constructing the effective metrics for the geometry and
studying the motion of light rays in these two space-times, we
establish the criteria necessary to distinguish the two types of
charges from the motion of photons. We show that it is not
possible to distinguish the two types of charges only from the
motion of light rays around the black hole. The paper is organized
as follows: In Sec.~\ref{sec-basiceq} we briefly review the main
equations to construct black hole solutions in general relativity
coupled to nonlinear electrodynamics. In Sec.~\ref{sec-effective}
we derive the effective metrics of the electrically and
magnetically charged black holes and in Sec.~\ref{sec-light-prop}
we study the motion of light rays in these space-times. Finally,
in Sec. \ref{sec-discussion} we discuss and summarize results.
Throughout the paper, we adopt the following signature convention
$(-,+,+,+)$ for the space-time metric and make use of natural
units, thus setting $c=\hbar=G=1$.

\section{Basic equations}\label{sec-basiceq}

A generic theory of general relativity coupled to nonlinear
electrodynamics is characterized by the action
\bear\label{action}
S = \int d^4x \sqrt{-g} \cal{L}\ ,
\ear
with the Lagrangian density given by
\bear\label{lagrangian}
{\cal{L}}=\frac{1}{16\pi}\left[R-\L(F)\right],
\ear
where $g$ and $R$ are the determinant of the metric tensor and the
Ricci scalar, respectively while $\L$ is the Lagrangian density
describing the nonlinear electrodynamics theory, which is a
function of the Faraday tensor $F^{\mu\nu}$ through $F\equiv
F_{\mu\nu}F^{\mu\nu}$. The electromagnetic field tensor satisfies
$F_{\mu\nu}=\partial_\mu A_\nu-\partial_\nu A_\mu$, with $A^\mu$
being the 4-potential. Since the Faraday tensor, $F_{\mu\nu}$, is
antisymmetric, it has only six nonvanishing components: i.e. three
for the electric field and three for the magnetic field.

To construct a solution of the theory described by~(\ref{action}),
one needs to solve Einstein's field equations given by
\bear\label{Einstein-eq}
G_{\mu\nu}=T_{\mu\nu}\ ,
\ear
where $G_{\mu\nu}$ is the Einstein tensor and $T_{\mu\nu}$ is the
energy-momentum tensor of the nonlinear electrodynamics field.
Note that, for simplicity, in equation (\ref{Einstein-eq}) the
coefficient $8\pi$ has been absorbed in the energy-momentum
tensor. The energy-momentum tensor of nonlinear electrodynamics is
given by
\bear\label{em-tensor}
T_{\mu\nu}=2\left(\L_F F_\mu^{\ \alpha} F_{\nu\alpha}-\frac{1}{4}g_{\mu\nu}\L\right)\ ,
\ear
where $\L_F=\partial_F\L$. At the same time, the electromagnetic
field is governed by Maxwell's equations for nonlinear electrodynamics, which can be written as
\bear\label{Maxwell-eq} \nabla_\nu\left(\L_F F^{\mu\nu}\right)=0\
, \quad \nabla_\nu {}^\ast F^{\mu\nu}=0
\ear
Where $\ast F^{\mu\nu}=\varepsilon^{\mu\nu\alpha\beta}
F_{\alpha\beta}/2$ is the dual electromagnetic strength tensor.
The electromagnetic 4-potential can be written in spherical
coordinates $\{t,r,\theta,\phi\}$ in the following form:
\bear\label{ansatz}
A_\mu=\varphi(r)\delta_\mu^t-Q_m\cos\theta\delta_\mu^\phi\ ,
\ear
where $\varphi(r)$ and $Q_m$ are the electric potential and total
magnetic charge, respectively. The exterior of spherically
symmetric, static, electrically and magnetically charged compact
objects is described by the same line elements which can be
written in general as
\bear\label{line-element}
ds^2=-f(r)dt^2+\frac{dr^2}{f(r)}+r^2d\Omega^2\ ,
\ear
where $d\Omega^2=d\theta^2+\sin^2\theta d\phi^2$ is the metric on
the unit two-sphere and the metric function $f(r)$ is given in the parameterized form as
\bear\label{metric-function}
f(r) = 1 - \frac{2m (r)}{r}\ ,
\ear
with the mass function, $m(r)$ determined by the Lagrangian
density of the nonlinear electrodynamics. In the absence of the
electromagnetic field, the mass function takes constant value
$m(r)=M$, consistent with the description of a purely
gravitational mass. For the sake of our further calculations, here
below we will briefly review the main points of the formalism for
deriving electrically and magnetically charged black hole
solutions.

\subsection{Electrically charged solution}

If the space-time is electrically charged, then the 4-potential of
the electromagnetic field is given solely by the first term in
equation (\ref{ansatz}), as $A_t=\varphi(r)$. The electromagnetic
field 2-form can be written as
\bear\label{em2-form-el}
\textbf{F}_2=\varphi'(r)\textbf{d}r\wedge\textbf{d}t\ .
\ear
Note that $F=-2\varphi'^2$. By solving the non linear Maxwell's
equations~(\ref{Maxwell-eq}), we arrive at the expression for the
total electric charge inside a sphere with radius $r$
\bear\label{total-charge}
Q_e=r^2 \L_F \varphi'\ .
\ear
At this point, to construct a solution, we need to solve
Einstein's equations~(\ref{Einstein-eq}), which, for this system,
reduce to only two independent equations. By solving them, we
obtain the following:
\bear
&&\L=\frac{2m''}{r},\label{lag-el1}\\
&&\L_F=\frac{2m'-rm''}{2r^2\varphi'^2}.\label{lag-el2}
\ear
By using (\ref{total-charge}) and (\ref{lag-el2}), we find the
expression for the scalar electric potential as
\bear\label{el-pot}
\varphi=\frac{3m-rm'}{2Q_e}\ .
\ear
If the mass function is constant, then, we immediately recover the
Schwarzschild solution.  If the electromagnetic field is linear,
i.e., the Lagrangian density is linear function of $F$, by solving
eqs. (\ref{lag-el1}), (\ref{lag-el2}), and (\ref{el-pot}), one can
find the Reissner-Nordstr\"{o}m solution with mass function
$m(r)=M-Q_e^2/2r$, and the Coulomb potential $\varphi\sim Q_e/r$.

\subsection{Magnetically charged solution}

If the black hole is magnetically charged, then the 4-potential of
the electromagnetic field is given by the second term in
equation~(\ref{ansatz}), as $A_\phi=-Q_m\cos\theta$. The
electromagnetic field 2-form can then be written as
\bear\label{em2-form-el}
\textbf{F}_2=Q_m\sin\theta\textbf{d}\theta\wedge\textbf{d}\phi\ .
\ear
Note the electromagnetic field strength is $F=2Q_m^2/r^4$. By
solving Einstein's equations~(\ref{Einstein-eq}), we obtain the
following expressions for the Lagrangian density:
\bear
&&\L=\frac{4m'}{r^2},\label{lag-mg1}\\
&&\L_F=\frac{r^2(2m'-rm'')}{2Q_m^2}.\label{lag-mg2}
\ear
In the case that the electromagnetic field is linear, i.e. for
Maxwell's theory, we have that $\L=F$ and by solving the above
equations, we arrive at the mass function $m=M-Q_m^2/2r$ that
again represents the Reissner-Nordstr\"{o}m space-time with a
magnetic charge. We see that in the case of linear electrodynamics
the two charges are not distinguishable in
Reissner-Nordstr\"{o}m's solution. On the other hand, in the
nonlinear theory the two cases produce two different effective
geometries. We shall now investigate whether such space-times may
be distinguished by looking at the trajectories of light rays.

\section{Effective metrics}\label{sec-effective}

As we mentioned in the case where the line
element~(\ref{line-element}) is a solution of the field equations
for general relativity coupled to nonlinear electrodynamics,
light rays do not propagate along the null geodesics of the
space-time metric, instead, they follow the null geodesics of the
effective metric obtained from the metric tensor
\cite{Novello:PRD:2000,Bronnikov:PRD:2001}
\bear\label{tensor-eff-metric}
g_{eff}^{\mu\nu}=\L_F g^{\mu\nu}
-\L_{FF}F_\lambda^{\ \mu} F^{\lambda\nu}\ .
\ear

\subsection{Electrically charged case}

In the electrically charged case the effective metric is written
in the following form:
\bear\label{eff-metric-el}
ds^2=-\frac{f(r)}{\Phi}dt^2+\frac{1}{\Phi f(r)}dr^2
+\frac{r^2}{\L_F}d\Omega^2\ ,
\ear
where $\Phi=\L_F+2F\L_{FF}$. When written in terms of the mass
function $m(r)$, then the above line element takes the following
form:
\bear\label{eff-metric-el1}
ds^2&=&-\frac{r^2 (r-2 m) \left(r
m^{'''}-m''\right)}{4 Q_e^2}dt^2+ \\ \nonumber
&&+\frac{r^4\left(r
m^{'''}-m''\right)}{4 Q_e^2 (r-2 m)}dr^2+\frac{r^4 \left(2 m'-r
m''\right)}{2 Q_e^2}d\Omega^2\ .
\ear

\subsection{Magnetically charged case}

On the other hand, the effective metric for a magnetically
charged black hole is written in the following form:
\bear\label{eff-metric-mg}
ds^2=-\frac{f(r)}{\L_F}dt^2+\frac{1}{\L_F f(r)}dr^2
+\frac{r^2}{\Phi}d\Omega^2\ , \ear
where again $\Phi=\L_F+2F\L_{FF}$. And again, when written in
terms of the mass function $m(r)$, it takes the following form:
\bear\label{eff-metric-mg1}
ds^2&=&-\frac{2 Q_m^2 (r-2 m)}{r^3
\left(2 m'-r m''\right)}dt^2+ \\ \nonumber
&&+\frac{2Q_m^2}{r(r-2m)\left(2m'-r
m''\right)}dr^2+\frac{4Q_m^2}{r^2 m^{'''}-r m''}d\Omega^2\ .
\ear
We see that for a generic mass function $m(r)$ the effective
metric of an electrically charged black
hole~(\ref{eff-metric-el1}) differs from that of a magnetically
charged one~(\ref{eff-metric-mg1}).

\section{Light rings and gravitational lensing}\label{sec-light-prop}

Before finding the equations governing the motion of light rays
in the effective space-time metrics for the electrically and
magnetically charged cases, one may write the effective
metrics~\eqref{eff-metric-el1} and~\eqref{eff-metric-mg1} in the
following unified form:
\bear\label{gen-line-element}
ds^2=-A(r)dt^2+B(r)dr^2+C(r)d\Omega^2\ .
\ear
Taking into account the symmetry of the space-time, one can easily
notice  that the momenta $p^\mu$ corresponding to time, $t$, and
azimuthal angle, $\phi$, are conserved. These are related to the
energy, ${\rm E}$, and angular momentum, ${\rm L}$, of test
particles and photons. Restricting the attention to motion in the
equatorial plane, $\theta=\pi/2$, the conserved quantities are
given by
\bear\label{eq-motion1}
{\rm E}=A\dot{t}\ , \qquad {\rm L}=C\dot{\phi}\ .
\ear
Since for photon's motion we have $p^\mu p_\mu=0$, the radial
component of the 4-velocity of light rays can be written in terms
of the conserved quantities as
\bear\label{rad-vel}
\dot{r}^2=\frac{1}{AB}\left({\rm E^2- V_{eff}}\right),
\quad {\rm with} \quad {\rm V_{eff}=L^2}\frac{A}{C}\ .
\ear
Then circular null geodesics are obtained by imposing
$\dot{r}=0=\ddot{r}$. Therefore, setting to zero the expression
inside parenthesis in (\ref{rad-vel}), one can find that the
energy for photons on circular orbits, while the radius of
circular null geodesics i.e., the light ring, is determined by
the radius for which $\ddot{r}=0$, corresponding to the solution
of the following equation:
\bear\label{lightring}
AC'-A'C=0\ .
\ear

Before turning to the further relativistic effects, let us
consider the effective potentials in terms of the effective
metrics of the electrically and magnetically charged black holes
which are given by equations~\eqref{eff-metric-el1} and
\eqref{eff-metric-mg1}.

In the case of the Reissner-Nordstr\"{o}m black hole with mass
function $m=M-Q^2/2r$, we recover the effective potential
$V_{eff}=(1-2M/r+Q^2/r^2){\rm L}^2/r^2$. As we have mentioned in
section \ref{sec-basiceq}, for linear electrodynamics, the
Reissner-Nordstr\"{o}m solution describes both electrically as
well as magnetically charged black hole space-times via the same
line element~(\ref{line-element}). In other words, the mass
functions for electrically and magnetically charged black holes in
linear electrodynamics coincide. From
equations~\eqref{eff-metric-el1} and \eqref{eff-metric-mg1} it is
easy to realize that even though the effective metrics differ, the
effective potentials for the motion of light rays in the
electrically and magnetically charged cases are the same. In fact,
the effective potentials from equation~\eqref{rad-vel} in terms of
the effective metrics of the electrically~\eqref{eff-metric-el1}
and magnetically \eqref{eff-metric-mg1} charged black holes take
the form
\bear\label{eff-pot-el}
{\rm V_{eff}}=\frac{f(m''-r
m^{'''})}{2 r \left(r m''-2 m'\right)}{\rm L}^2\ .
\ear
Consequently, the radii of the light rings~(\ref{lightring}) are
identical and it is not possible to distinguish the electrically
charged from the magnetically charged case solely from the
location of the photon sphere.

Despite the above result that the expression inside the
parenthesis in equation~\eqref{rad-vel} is identical for both
electrically and magnetically charged cases, the product of the
metric functions $A$ and $B$ differs in the two cases which
suggests that there may be other ways of distinguishing the two
charges from the motion of photons. In fact in the electrically
charged case we get
\bear
AB=\frac{r^6 \left(m''-r m^{'''}\right)^2}{16 Q_e^4}\ ,
\ear
while for the magnetically charged case we have
\bear
AB=\frac{4 Q_m^4}{r^4 \left(r m''-2 m'\right)^2}\ .
\ear
In principle, considering other relativistic effects where these
terms play a role may allow to distinguish the type of charge of
the black hole. With this objective in mind, below we will
consider gravitational lensing in the strong field regime of both
space-times. As it was mentioned before, even though electrically
and magnetically charged black holes in nonlinear electrodynamics
are described by the same line element, their effective metrics
differ and light rays follow the null geodesics of the effective
space-time. Let us then consider light rays passing close to a
compact, massive object and evaluate how light rays deviate from a
straight trajectory while following the geodesics of the
space-time surrounding a massive compact object.

We first derive the equation for the deflection angle of
gravitational lensing in a generic spherically symmetric
space-time given by equation~(\ref{gen-line-element}). By using
the equations of motion~(\ref{eq-motion1}) and (\ref{rad-vel}),
we find the deflection angle of the light ray from
\bear\label{def-angle}
\frac{d\phi}{dr}=\frac{\sqrt{B}}{\sqrt{C}
\sqrt{\frac{A_0}{C_0}\frac{C}{A}-1}}\ ,
\ear
where we have denoted the light ring solution of
equation~\eqref{lightring} with $r_0$ and consequently quantities
$X$ evaluated at $r_0$ are indicated via the subscript $X_0$. One
can easily notice that as the radius $r$ tends to $r_0$
($r\rightarrow r_0$), the deflection angle diverges~\footnote{The
deflection angle of light rays is determined by solving the
following integral: \bear \nonumber
\Delta\phi(r_0)=2\int_{r_0}^\infty
dr\left(\frac{d\phi}{dr}\right)-\pi\ . \ear }. This fact shows
that at $r_0$ light rays move along circular orbits.

Another interesting relativistic effect associated with
gravitational lensing is the apparent time delay in the
propagation of light rays passing near a massive object, i.e. the
delay in travel time of the light ray from the source to the
receiver. The time delay is found by using the equations of
motion~(\ref{eq-motion1}) and (\ref{rad-vel}) and it can be
written as
\bear\label{time-delay}
\frac{dt}{dr}=\frac{\sqrt{B}}{\sqrt{A}\sqrt{1-\frac{C_0}{A_0}\frac{A}{C}}}\ ,
\ear
From equation~\eqref{time-delay} we can see that when the turning
point of the light ray reaches the light ring of the black hole at
radius $r_0$, the delaying time diverges and the light ray never
reaches the observer.

Now let us evaluate the deflection of light rays and the time
delay due to the gravitational lensing for a central object that
is either electrically or magnetically charged.

\subsection{Electrically charged case}

Rewriting the expressions (\ref{def-angle}) and (\ref{time-delay})
in terms of the effective metrics of the electrically charged
black hole (\ref{eff-metric-el1}) we obtain
\bear\label{def-angle-el}
\frac{d\phi}{dr}&=&\sqrt{\frac{z}{2f}}\frac{1}{\sqrt{\frac{f_0z_0}{fz}-1}}\ , \\
\label{time-delay-el}
\frac{dt}{dr}&=&\frac{1}{f\sqrt{1-\frac{fz}{f_0z_0}}}\ ,
\ear
where
\bear
z=\frac{m''-rm^{'''}}{r \left(r m''-2 m'\right)}\ ,\nonumber
\ear
and the zero subscript indicates the value of a function evaluated
at the light ring $r_0$.

\subsection{Magnetically charged case}

On the other hand, in terms of the effective metrics of the
magnetically charged black hole (\ref{eff-metric-mg1}),
expressions (\ref{def-angle}) and (\ref{time-delay}) take the
following form:
\bear\label{def-angle-mg}
\frac{d\phi}{dr}&=&\sqrt{\frac{z}{2f}}\frac{1}{\sqrt{\frac{f_0z_0}{fz}-1}}\ , \\
\label{time-delay-mg}
\frac{dt}{dr}&=&\frac{1}{f\sqrt{1-\frac{fz}{f_0z_0}}}\ ,
\ear

Interestingly, despite the fact that the equations of motion
governing the propagation of light rays in the electrically and
magnetically charged black hole space-times are different, the
trajectories followed by light rays near a massive compact object
coincide. Therefore, we conclude that it is not possible to
distinguish an electrically charged black hole from a magnetically
charged one just by measuring the deflection of light rays.

\section{Discussion}\label{sec-discussion}

We have shown that, despite the fact that the effective metrics of
electrically and magnetically charged black holes in general
relativity coupled to nonlinear electrodynamics are different,
photons follow the same trajectories and therefore, the
observation of light propagation alone can not distinguish the two
kinds of charges. Therefore, to gain some further insights it may
be worth to return to the field equations presented in section
\ref{sec-basiceq} and analyze their dependence on the non linear
electromagnetic Lagrangian term. In the electrically charged case,
if $\L_F$ in (\ref{lag-el2}) is written in terms of the total
electrical charge of the black hole (\ref{total-charge}), and
compared to the one (\ref{lag-mg2}) for the magnetically charged
case, one would easily notice the following relation:
\bear
\left(\L_F\right)_{e}=\frac{1}{\left(\L_F\right)_{m}}\ .
\ear
Moreover, it is easy to show that the following relation between
the electromagnetic field strengths in both cases holds:
\bear
\left(F \L_F^2\right)_{e}=-F_m\ .
\ear
If the effective metrics in equations~\eqref{eff-metric-el} and
\eqref{eff-metric-mg} for both cases are rewritten taking into
account the relations given above and considering the fact that
the conformal factor does not affect the motion of light
rays~\cite{ToshmatovPRD:96}, one can easily obtain the following
relation:
\bear
\left(\frac{\L_F}{\Phi}\right)_e=\left(\frac{\Phi}{\L_F}\right)_m\ ,
\ear
and consequently, end up with the identical space-times.

Based on the fact that the light ray does not follow null
geodesics in general relativity coupled to nonlinear
electrodynamics
\cite{Novello:PRD:2000,Novello:PRD:2001,ObukhovPRD:66}, instead
they follow the null geodesics of the effective metric, in this
paper we have studied the possibility that the type of charge with
which the black hole space-time in endowed may be distinguished
from the motion of light rays in the equatorial plane of a black
hole geometry. To this aim, we considered that a space-time metric
(\ref{line-element}) with metric function (\ref{metric-function})
that is a solution of the field equations of general relativity
minimally coupled to the nonlinear electrodynamics and describes a
static and spherically symmetric black hole. Then, we considered
separately the two cases in which the black hole is endowed with
an electrical or a magnetic charge and studied the physical
effects related to the propagation of light rays. We have shown
that light rays follow the same trajectory in both cases, despite
of the fact that the corresponding effective metrics are
different. Therefore the observation of photon trajectories alone
is not able to distinguish the two kinds of charges.

\section*{Acknowledgements}

BT and BA acknowledge the support of Ministry of Innovative
Development of the Republic of Uzbekistan Grants No.~VA-FA-F-2-008
and No.~MRB-AN2019-29. DM is supported by the Ministry of
Education of the Republic of Kazakhstan's target program
IRN:~BR05236454 and Nazarbayev University Faculty Development
Competitive Research Grant No.~090118FD5348.

\label{lastpage}

\bibliography{Toshmatov_references}

\end{document}